# FINANCIAL MARKETS, FINANCIAL INSTITUTIONS AND INTERNATIONAL TRADE: EXAMINING THE CAUSAL LINKS FOR INDIAN ECONOMY

Ummuhabeeba Chaliyan

Doctoral Candidate,

Department of Economics and Finance,

Birla Institute of Technology and Science Pilani Hyderabad campus

(Email: p20180020@hyderabad.bits-pilani.ac.in)

Mini P. Thomas

Assistant Professor,

Department of Economics and Finance,

Birla Institute of Technology and Science Pilani Hyderabad campus

(Email: mini@hyderabad.bits-pilani.ac.in)

December 3, 2021

# FINANCIAL MARKETS, FINANCIAL INSTITUTIONS AND INTERNATIONAL TRADE: EXAMINING THE CAUSAL LINKS FOR INDIAN ECONOMY

Ummuhabeeba Chaliyan        Mini P. Thomas

## Abstract

This study investigates whether a uni-directional or bi-directional causal relationship exists between financial development and international trade for Indian economy, during the time period from 1980 to 2019. The empirical analysis utilizes three measures of financial development created by IMF, namely, financial institutional development index, financial market development index and a composite index of financial development, encompassing dimensions of financial access, depth and efficiency. Johansen cointegration, vector error correction model and vector auto regressive model are estimated to examine the long run relationship and short run dynamics among the variables of interest. The econometric results indicate that there is indeed a long run causal relationship between the composite index of financial development and trade openness. Cointegration is also found to exist between trade openness and index of financial market development. However, there is no evidence of cointegration between financial institutional development and trade openness. Granger causality test results indicate the presence of uni-directional causality running from composite index of financial development to trade openness. Financial market development is also found to Granger cause trade openness. In contrast, trade openness is found to promote financial institutional development in the short run. Empirical evidence thus underlines the importance of formulating policies which recognize the role of well-developed financial markets in accelerating international trade of Indian economy.

**Key words:** Financial Development, Financial Markets, Financial Institutions, Trade openness, Cointegration, VEC, VAR

**JEL Classification:** F14, G10, G20, C01, C22

**Introduction**

Financial development and international trade are key drivers of the economic growth performance of a country. India has a strongly diversified financial sector comprising commercial banks, insurance companies, non-banking financial companies, cooperatives, pension funds and mutual funds. Financial sector reforms initiated in India in the 1990s, based on the report by Narasimhan Committee, mainly focused on banking system and capital market. The reforms aimed at putting an end to the then prevailing regime of financial repression, enable price discovery by the market determination of interest rates, and maintain financial stability even in the face of domestic or external shocks (RBI 2007). Financial development, measured as domestic credit to private sector as percentage of GDP, rose from 20.5 in 1980 to 50.15 in 2019 for Indian economy. It peaked at 52.4 percent in 2013. Interestingly, banks have provided more than 50 percent of the domestic credit requirement of private sector in India since 2010 (World Bank 2019). The state of development of financial sector varies across countries. Consequently, there has been rigorous debate about the desirable mode of financial system. At the same time, international trade liberalization achieved through the adoption of several trade liberalization policies in different years. An important wave in this respect was the implementation of 5 years Export Import (EXIM) policy in 1992.For India, Trade openness to GDP ratio stood at 15.4 percent in 1980, after which it steadily rose and attained a peak of 55.8 percent in 2012. It has exhibited a declining trend ever since and stood to 40.01 in 2019 (World Bank 2019).

Financial development is increasingly becoming an important contributor to the economic growth of countries. Seminal work by (Goldsmith 1969), (McKinnon 1973) and (Shaw 1973) have analyzed the role of financial intermediaries in boosting long run growth of economies. Other theoretical studies also contributed to the debate in the successive years, which include (Levine 1997, 702, Levine and Zervos 1998, 550); (Levine, Loayza, and Beck 2000, 40) and (Luintel and Musahid 1999, 399). Another work (Murinde and Eng 1994, 400) is examining the country specific study about Singapore for the period 1979 - 1999, finds the causality from financial development to economic growth. Evidence obtained from the study in Turkey for the period 1989- 2007, supports the bidirectional causality exists between financial development, trade openness and growth (Yucel 2009).

Studies which examined the role of financial development in economic growth of countries are numerous and well established in the literature, compared to studies which unravel the link between financial development and international trade. The present study aims to examine whether a unidirectional or bi-directional relationship exists between financial development and international trade in case of Indian economy, by using time series techniques. This study brings in a new dimension to existing studies on this subject, by resorting to the broad based measure of financial development created by IMF, to account for all possible variations in the financial system. Since financial development is a complex multidimensional variable, the typical proxies such as ratio of private credit to GDP or stock market capitalization can't fully account for the concept. Broad financial development index of IMF encompasses three dimensions of efficiency, access and depth, with respect to both financial institutions and financial markets (Svirydzenka 2016).

## Literature Review

The Heckscher Ohlin and Vanek theory of international trade states that factor content of a commodity is most important for trade. It postulates that a country well-endowed with a particular factor will engage in the production and trade of that commodity which intensively use the abundant factor. (Kletzer and Bardhan 1987, 65) was the first study to link credit markets to international trade patterns. They postulated that countries with identical technology and endowments may still differ in terms of comparative cost advantages, because of credit market imperfections, arising from moral hazard considerations and asymmetric information. Path breaking work by (Beck 2002, 120) theoretically modeled the role of financial intermediaries in boosting large scale, high return projects, and proved that countries with a higher state of financial development have a comparative advantage in manufacturing. The model was validated using a panel dataset for 30 years for 65 countries. Trade in manufactured goods was specified as a function of private credit as a share of GDP and other control variables such as initial level of real per capita GDP, black market premium, real per capita capital, population and growth rate of terms of trade. The estimation was carried out using Ordinary Least squares and Instrumental Variable Technique. After controlling for country specific effects and possible reverse causality, the study found that financial development does exert a significant causal impact on two measures of international trade, namely the level of exports and the trade balance of manufactured goods.

Few studies (Beck 2002, 120) and (Vlachos and Svalryd 2005, 114) have analyzed the link between financial development and international trade, from the point of view of economies of scale. They found that financial sector can facilitate trade immensely. A well developed financial sector can source savings to private sector and assist entrepreneurs to engage in more business activities, thereby overcoming credit constraints. Since manufacturing sector exploits increasing economies of scale, it can reap higher profit coupled with the high level of financial development associated with manufacturing sector. In every country, the sector which faces demand shock has to protect from risk. This implies that international trade patterns are highly dependent on differences in financial development, with a highly developed financial system permitting the country to specialize in risky goods (Baldwin 1989, 145). Availability of trade credit is determined by the level of financial health and width of network of issuing banks. These factors can positively contribute to the flexibility and liquidity of funding and supply of trade credit (Jain, Gajbhiye, and Tewari 2019).

(Do and Levchenko 2004) built a theoretical model wherein international trade boosts growth of financially dependent sectors and financial system in a wealthy country. The poor country begins to import the financially dependent good from the rich country rather than produce it domestically, implying a decline of the domestic financial system and demand for external finance. The model was empirically validated using a panel dataset of 77 countries, from 1965 to 1995. Financial development was specified as a function of trade openness and control variables such as initial level of private credit to GDP, initial per capita GDP, a measure of human capital (average years of secondary schooling in the population), as well as legal origin dummies. Financial development was measured using three alternative indicators, namely, the ratio of private credit to GDP, the ratio of liquid liabilities to GDP, and claims of deposit money banks on nonfinancial domestic sectors as share of GDP. This study thus provides theoretical and empirical basis for direction of causality running from trade openness to financial development.

As trade finance is an important factor determining flow of imports and exports, financial development is always positively correlated with volume of trade (Liston and McNeil 2013, 13). Studies carried out using multi industry approach found that financial development and ratio of export to domestic sales is positively associated in capital intensive industries, whereas the trade share is much lower in labour intensive industries. But the net effect can be offset at the aggregate level (Leibovici 2018). Empirical studies indicate positive short run and negative long run association between financial development and international trade, along with unidirectional causality running from financial development to international trade (Bilas, Bosnjak, and Novak 2017, 84). Economic growth is linked to international trade through the medium of financial development, whereby economic growth increases financial development and thus enhances trade participation of countries (Kar, Nazlioglu, and Agir. 2013, 139).

Countries that trade manufactured commodities can elevate the financial system of that country because such economies demand more external finance, which leads to greater financial development of country (Samba and Yu 2009, 65). Export oriented firms demand more external finance to meet high fixed cost. Therefore, destabilizing financial conditions affect export oriented firms more than it affects domestic oriented ones (Feng and Lin2013, 44). Another interesting inter-relationship between financial development and international trade is found via the linkage between imports and debt financed consumption. Higher domestic demand is financed by inflow of foreign loans. Both the appreciation of domestic currency due to higher imports and inflow of foreign loans resulted in current account deficit in European transition countries (Aristovnik 2008; Zakharova 2008; Bakker and Gulde 2010, 120). When developing and developed countries were analyzed separately for the period between 1961 and 2020 to assess the linkage between financial development and international trade, financial development was found to be a key factor which promotes trade participation of countries. The direction of causality varied among countries, depending on their level of economic development (Kiendrebeogo 2012).

Financial development (private credit and money supply) has a positive influence on trade volume by increasing productivity and technology up gradation through efficient allocation of financial resources (Kaushal and Pathak 2015, 10). Another empirical study investigated the relationship between exports, financial development and GDP growth in Pakistan by applying the Bound testing approach to cointegration and vector error correction model (VECM) based Granger causality test, and found the existence of the long run relationship among the variables (Shabaz and Rahman 2014, 164). Johansen multivariate approach to cointegration found no significant long run relation among economic growth, financial development and international trade for Nigeria during the time period from 1970 to 2005 (Chimobi 2010). (Arora and Mukherjee 2020, 285) examined the nexus between financial development and trade performance of Indian Economy using annual data from 1980 to 2016. The study resorted to the time series technique of the Auto regressive distributed lag bound testing approach to check for the presence of cointegration. Financial development was measured using the widely used indicator of private credit as a percentage of GDP. International trade performance was measured using three alternative indicators, namely, manufactured exports as a percentage of GDP, manufactured imports as a percentage of GDP and net manufactured exports as a percentage of GDP. International trade in primary goods and services were not covered in this study. They found the existence of a long run equilibrium relationship between international trade in manufactured goods and financial development and existence of unidirectional causality from financial development to net exports of manufactured goods. Empirical evidence indicated absence of reverse causality from international trade to financial development in case of Indian economy.

Based on the above literature review, it is found that that there is no conclusive evidence with regard to the direction of causality between the state of financial development and international trade of a country. The present study is trying to contribute to this growing literature by analyzing the linkage between financial development and international trade specifically for Indian economy. There is a dearth of Indian studies which have examined the relationship between financial development and international trade in aggregate (inclusive of agriculture, manufacturing and services). This paper also adds a new dimension to the existing studies by measuring financial development using the broad based index created by IMF. This indicator is definitely superior to ratio of private credit to GDP, as a proxy for financial development. This study also throws light on the nature of relationship between trade openness and two of the sub-indices of IMF index of financial development, namely, financial market development and financial institutional development.

## Data and Methodology

The empirical model to estimate the relationship between financial development and international trade in India is specified as given in equations (1), (2) and (3).

TRADE = f (FD, LGDP, REER)                    (1)

TRADE = f (FID, LGDP, REER)                   (2)

TRADE = f (FMD, LGDP, REER)                   (3)

The term TRADE in the equations (1), (2) and (3) denote trade openness, measured as the sum of exports and imports of a country, expressed as a percentage of GDP. FD, FID and FMD in equations (1), (2), and (3) denote financial development, financial institutional development and financial market development indices respectively. The term LGDP in all the three equations represents log of GDP per capita and the term REER in all the three equations represent the real effective exchange rate. GDP and REER are taken as control variables in the model specification.

Annual data representing GDP per capita and trade openness are taken from World Development Indicators 2021 database of World Bank. Annual time series data on financial development, financial institutional development and financial market development are obtained from the Financial Development Index Database of IMF. Time series data on India's real effective exchange rate is taken from RBI's Handbook of statistics on Indian economy. The time span of the study ranges from 1980 to 2019.

The main aim of the paper is to test whether a one way or two way causal relationship exists between financial development and international trade for Indian economy. Time series econometric techniques are utilized for this purpose. Unit root tests are used to examine the stationarity properties and identify the order of integration of the above mentioned macroeconomic variables, since most of the economic and financial time series data tend to be non stationary. Second step of analysis involves testing for cointegration to check for the existence of a long run relationship amongst the variables of interest. Since all the variables are of the order of integration of one, Johansens cointegration technique (1988), which is based on maximum likelihood method, is most suitable for this study. After cointegration between the variables is established, the VEC Model is estimated to study the short run dynamics and direction of causality among the variables.

In the absence of cointegration for any of the equations (1), (2) and (3), VAR (Vector Autoregression) Model is estimated. The causality test is employed as any cointegrated system establishes an error correction mechanism that restricts the variable to deviate from its long run equilibrium. In applied econometric literature, the direction of causal relationship can be examined by using the well known Granger causality test (1988). In every step, the dependent variable is regressed on past values of itself and other independent variables. The optimum lag length for the model is chosen based on Akaike Information Criteria.

VECM representation for the equations 1 and 3 are as follows:

$$\Delta TRADE_t = \beta_0 + \sum_{j=1}^{m} \theta k \, \Delta TRADE_{t-j} + \sum_{j=1}^{m} \gamma k \, \Delta FD_{t-j} + \sum_{j=1}^{m} \tau k \, \Delta LGDP_{t-j} + \sum_{j=1}^{m} \varphi k \, \Delta REER_{t-j} + \lambda [TRADE_{t-1} - \alpha_0\hat{} - \alpha_1 \, FD_{-1} - \alpha_2\hat{} \, LGDP_{-1} - \alpha_3\hat{} \, REER_{-1}] + \varepsilon_t \qquad (4)$$

$$\Delta TRADE_t = \beta_1 + \sum_{j=1}^{m} \delta k \, \Delta TRADE_{t-j} + \sum_{j=1}^{m} \chi k \, \Delta FMD_{t-j} + \sum_{j=1}^{m} \sigma k \, \Delta LGDP_{t-j} + \sum_{j=1}^{m} \eta k \, \Delta REER_{t-j} + \rho [TRADE_{t-1} - \alpha_4\hat{} - \alpha_5 \, FMD_{-1} - \alpha_6\hat{} \, LGDP_{-1} - \alpha_7\hat{} \, REER_{-1}] + \pi_t \qquad (5)$$

Here, $\Delta$ indicates first difference operator, $\lambda$ and $\rho$ are representing speed of adjustment to attain long run equilibrium and $\varepsilon_t$ and $\pi_t$ are error terms.

Vector Auto regression framework for equation 2 is as follows:

$$TRADE_{i,t} = \Pi_{11} \, TRADE_{i,t-k} + \Pi_{12} \, FID_{i,t-k} + \Pi_{13} \, LGDP_{i,t-k} + \Pi_{14} \, REER_{i,t-k} + \varepsilon_{\delta i,t} \qquad (6)$$

$$FID_{i,t} = \Pi_{21} \, FID_{i,t-k} + \Pi_{22} \, TRADE_{i,t-k} + \Pi_{23} \, LGDP_{i,t-k} + \Pi_{24} \, REER_{i,t-k} + \varepsilon_{\tau i,t} \qquad (7)$$

$$LGDP_{i,t} = \Pi_{31} \, LGDP_{i,t-k} + \Pi_{32} \, TRADE_{i,t-k} + \Pi_{33} \, FID_{i,t-k} + \Pi_{34} \, REER_{i,t-k} + \varepsilon_{\gamma i,t} \qquad (8)$$

$$REER_{i,t} = \Pi_{41} \, REER_{i,t-k} + \Pi_{42} \, TRADE_{i,t-k} + \Pi_{43} \, LGDP_{i,t-k} + \Pi_{44} \, FID_{i,t-k} + \varepsilon_{\varphi i,t} \qquad (9)$$

Afterwards, regression diagnostic tests are carried out to check for serial autocorrelation, multicollinearity, heteroscedasticity, normality of residuals and specification error. Breusch Godfrey LM Test, Variance Inflation Factor, White heteroskedasticity test and Jarque-Bera test and RAMSEY RESET Test are implemented to ensure that the estimated model does not suffer from the above problems. The presence of structural breaks in the model is detected with the help of a sequential Bai-Perron test (2003).

## Empirical Results

Countries across the world are experiencing greater integration of their domestic economies with the world economy. Economic growth of a country is influenced to a great extent by the growth of its financial sector along with other factors such as international trading environment.

Figure 1: Growth trend of Broad based Index of Financial Development

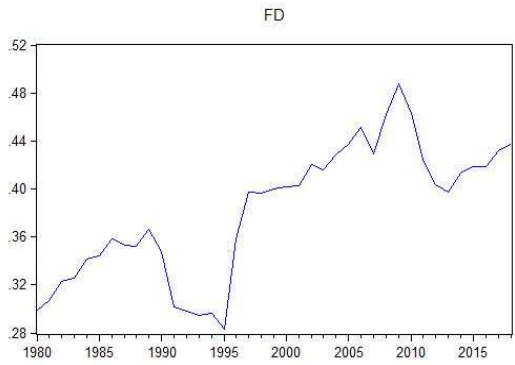

Source: IMF, 2021

Figure 2: Growth Trend of Financial Institutional Development

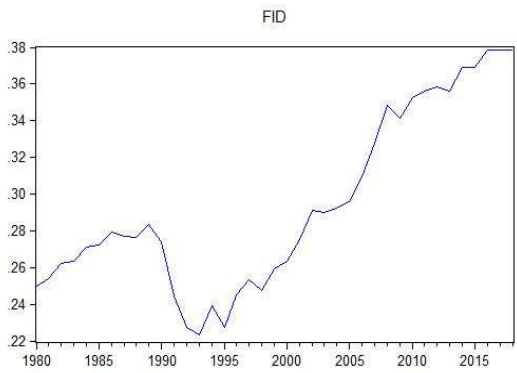

Source: IMF, 2021

Figure 3: Growth Trend of Financial Market Development

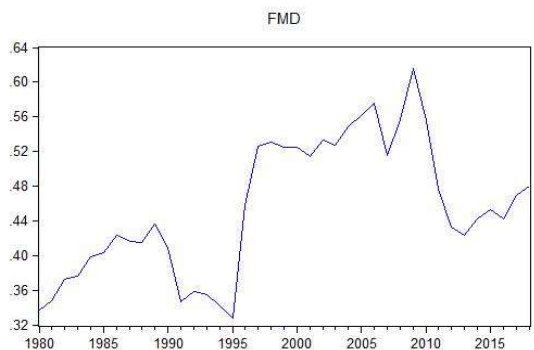

Source: IMF, 2021

Figure 1 plots the growth trend of the broad-based index of financial development created by IMF. Figures 2 and 3 are showing the growth trend of the sub-indices of financial market development and financial institutional development of IMF. While comparing the graphical plots, it is found that the growth trend of the financial development index exhibits the exact same pattern as the financial market development index. Financial institutional development is found to register a slightly different growth trajectory compared to the other two indicators.

In order to examine the long term relationship between the two macroeconomic variables of interest, it is important to test for stationarity and order of integration. Hence the standard unit root tests such as ADF (Dickey and Fuller 1979) and KPSS tests (Kwiatkowski, Phillips, Schmidt and Shin 1992) are initially carried out (see Table A1, Appendix). Since the paper is considering macroeconomic variables, it is wise to use a stationarity test which allows for unknown structural breaks as it avoids the possibility of spurious regression. The sequential Bai-Perron (2003) test claims for the presence of structural breaks in the series (see Table A3, Appendix). In the current study, Perron test (1997) unit root which allows for endogenous structural breaks is used in order to identify whether the series is stationary or not. Table 1 shows the results of unit root test based on Perron for all the variables included in the model. The null hypothesis of the Perron test is the presence of unit root in the series. Result implies that none of the variables are stationary at level but the first difference of these variables are stationary, that is all the variables are integrated of order one i.e, I(1).

Table 1: Perron unit root tests results.

| Variables | Levels | | | 1st Differences | | |
|---|---|---|---|---|---|---|
| | Test statistic | Critical value | Remarks | Test statistic | Critical value | Remarks |
| TRADE | -2.52 | -5.23 | I(1) | -6.75 | -5.23 | I(0) |
| FD | -4.28 | -5.23 | I(1) | -5.45 | -5.23 | I(0) |
| FID | -5.20 | -5.23 | I(1) | -5.98 | -5.23 | I(0) |
| FMD | -5.22 | -5.23 | I(1) | -5.34 | -5.23 | I(0) |
| LGDP | -2.52 | -5.23 | I(1) | -5.28 | -5.23 | I(0) |
| REER | -3.17 | -5.23 | I(1) | -7.01 | -5.23 | I(0) |

Source: Authors estimation. Note: Unit root tests are carried out at 5% significance level.

Different lag selection criteria for the three models to be estimated, are reported in Tables 2, 3 and 4. This study uses Akaike Information Criteria for all three equations, which is in support of optimum lag length of 5 for equations (1) and (3) and lag length of 2 for equation (2).

Table 2: Lag selection criteria result for equation (1)

| Lag | Log L | LR | FPE | AIC | SC | HQ |
|---|---|---|---|---|---|---|
| 0 | -201.1273 | - | 2.043927 | 12.06631 | 12.24589 | 12.12755 |
| 1 | 2.998838 | 348.2152 | 3.22e-05 | 1.000068 | 1.897927* | 1.306264 |
| 2 | 22.20845 | 28.24943 | 2.79e-05 | 0.811268 | 2.427414 | 1.362420 |
| 3 | 41.47593 | 23.80100 | 2.57e-05 | 0.619063 | 2.953497 | 1.415172 |
| 4 | 55.40166 | 13.92573 | 3.66e-05 | 0.741079 | 3.793800 | 1.782144 |
| 5 | 99.11271 | 33.42610* | 1.11e-05* | -0.888983* | 2.882025 | 0.397039* |

Source: Authors estimation. *indicates lag order selected by the criterion

Table 3: Lag selection criteria for equation (2)

| Lag | Log L | LR | FPE | AIC | SC | HQ |
|---|---|---|---|---|---|---|
| 0 | -180.2458 | - | 0.327727 | 10.23588 | 10.41182 | 10.29729 |
| 1 | 29.84445 | 361.8221 | 6.85e-06 | -0.546914 | 0.332819* | -0.239864* |
| 2 | 47.70304 | 26.78788* | 6.41e-06* | -0.650169* | 0.933350 | -0.097478 |
| 3 | 59.07589 | 14.53198 | 9.09e-06 | -0.393105 | 1.894200 | 0.405226 |
| 4 | 81.49289 | 22.16219 | 7.77e-06 | -0.771022 | 2.250797 | 0.272109 |

Source: Authors estimation. *indicates lag order selected by the criterion

Table 4: Lag selection criteria for equation (3)

| Lag | Log L | LR | FPE | AIC | SC | HQ |
|---|---|---|---|---|---|---|
| 0 | -222.0543 | - | 6.999739 | 13.29731 | 13.47689 | 13.35855 |
| 1 | -15.05446 | 353.1174 | 9.32e-05 | 2.062027 | 2.959886* | 2.368223 |
| 2 | 3.238808 | 26.90187 | 8.50e-05 | 1.927129 | 3.543275 | 2.478281 |
| 3 | 21.38408 | 22.41475 | 8.38e-05 | 1.800936 | 4.135370 | 2.597045 |
| 4 | 35.11510 | 13.73102 | 0.000121 | 1.934406 | 4.987127 | 2.975471 |
| 5 | 70.34910 | 26.94364* | 6.01e-05* | 0.802994* | 4.574003 | 2.089016* |

Source: Authors estimation. *indicates lag order selected by the criterion

Since all the macroeconomic variables are I(1), Johansen's cointegration approach can be used to check for the presence of long run equilibrium relationship between these variables. Once cointegration is established, VEC Model is estimated to study the short run dynamics. Results of Johansen cointegration test for all three equations (1), (2) and (3) are presented in Table 5, Table 6 and Table 7 respectively.

Table 5: Johansen Cointegration test results for Equation (1)

| Hypothesised No. of CE(s) | Trace Statistics | 0.05 Critical value | Max − Eigen Test Statistics | 0.05 Critical value |
|---|---|---|---|---|
| r = 0 | 109.8168* | 47.85613 | 54.79236* | 27.58434 |
| r <=1 | 55.02444* | 29.79707 | 42.17726* | 21.58434 |
| r <= 2 | 12.84718 | 15.49471 | 11.87843 | 14.26460 |
| r<=3 | 0.968748 | 3.841466 | 0.968748 | 3.841468 |

Source: Authors estimation. Note: *indicates 5% level of significance

Values of both trace and maximum eigen statistics in Table 5 are showing statistical significance. This indicates the existence of two cointegrating equations relating trade openness, composite index of financial development and other control variables at the 5% level of significance.

Table 6: Johansen Cointegration test results for Equation (2)

| Hypothesised No. of CE(s) | Trace Statistics | 0.05 Critical value | Max − Eigen Test Statistics | 0.05 Critical value |
|---|---|---|---|---|
| r = 0 | 40.89917 | 47.85613 | 23.59161 | 27.58434 |
| r <=1 | 17.30756 | 29.79707 | 12.46877 | 21.13162 |
| r <= 2 | 4.838794 | 15.49471 | 3.989456 | 14.26460 |
| r<=3 | 0.849338 | 3.841466 | 0.849338 | 3.841466 |

Source: Authors estimation. Note: *indicates 5% level of significance

On the basis of trace and max eigen values in Table 6, there is no evidence for cointegration between trade openness, index of financial institutional development and other control variables. Trace and maximum eigen value statistics in Table 7 indicate the presence of two cointegrating equations relating trade openness, financial market development and other control variables.

Table 7: Johansen Cointegration test results for Equation (3)

| Hypothesised No. of CE(s) | Trace Statistics | 0.05 Critical value | Max – Eigen Test Statistics | 0.05 Critical value |
|---|---|---|---|---|
| r = 0 | 117.3350* | 47.85613 | 63.58605* | 27.58434 |
| r <=1 | 53.74891* | 29.79707 | 40.97981* | 21.13162 |
| r <= 2 | 12.76910 | 15.49471 | 12.32449 | 14.26460 |
| r<=3 | 0.444604 | 3.841466 | 0.444604 | 3.841466 |

Source: Authors estimation. Note: *indicates 5% level of significance

Table 8: Short run coefficients based on VAR (2) framework

| | TRADE | FID | LGDP | REER |
|---|---|---|---|---|
| TRADE$_{-1}$ | 1.126 | 8.73E | 0.000102 | -0.3526 |
| SE | (0.165) | (0.00) | (0.00109) | (0.24287) |
| t statistics | [6.81]* | [0.017] | [0.0929] | [-1.452] |
| TRADE$_{-2}$ | -0.24 | 0.000756 | 0.000133 | 0.544 |
| SE | (0.17) | (0.0055) | (0.00118) | (0.2626) |
| t statistics | -1.34 | [1.36] | [0.112] | [2.07108]* |
| FID$_{-1}$ | -18.38 | 0.7584 | -0.0732 | 53.616 |
| SE | (57.37) | (0.177) | (0.379) | (84.27) |
| t statistics | [-0.32] | [4.26]* | [-0.192] | [0.636] |
| FID$_{-2}$ | 78.36 | -0.234 | -0.0309 | -179.77 |
| SE | (53.43) | (0.165) | (0.3533) | (78.48) |
| t statistics | [1.46] | [-1.41] | [-0.087] | [-2.29]* |
| LGDP$_{-1}$ | 40.411 | 0.229 | 1.0926 | 55.80 |
| SE | (30.95) | (0.095) | (0.204) | (45.46) |
| t statistics | [1.305] | [2.395]* | [5.338]* | [1.22] |
| LGDP$_{-2}$ | -48.32 | -0.-0.195 | -0.0622 | -43.49 |
| SE | (30.85) | (0.095) | (0.204) | (45.46) |
| t statistics | [-1.56] | [-2.04] | [-0.305] | [-0.95] |

| REER$_{-1}$ | 0.247 | 6.78E-05 | -0.0014 | 0.6550 |
|---|---|---|---|---|
| SE | (0.127) | (0.00040) | (0.0008) | (0.18) |
| t statistics | [1.939] | [0.171] | [-1.686] | [3.49]* |
| REER$_{-2}$ | -0.347 | 0.0004 | 0.0014 | 0.399 |
| SE | (0.144) | (0.00045) | (0.0009) | (0.212) |
| t statistics | [-2.404]* | [0.919] | [1.477] | [1.88] |
| Constant | 79.78 | -0.303 | -0.2604 | -109.47 |
| SE | (58.96) | (0.1828) | (0.389) | (86.615) |
| t statistics | [1.35] | [-1.658] | [-0.668] | [-1.26] |

Source: Authors estimation, *indicates 5% level of significance

Since VAR is an econometric tool used to analyze the dynamic impact of variables in the absence of cointegration, VAR is applied to equation (2). Table 8 presents the results of dynamic shocks of VAR relating to equations 6, 7, 8 and 9.

The above VAR model estimation results indicate that only trade openness with one year lag and real exchange rate with a two year lag exert a significant influence on current value of trade openness to GDP. Similarly, financial institutional development with one year lag and log of GDP per capita with one year lag is found to exert a statistically significant impact on the present value of financial institutional development. Trade openness with two year lag and financial institutional development with two year lag is also found to significantly impact upon current year's real exchange rate.

Short run adjustment parameters of equation (4) and (5) based on VECM are estimated and associated results are reported in Tables 9 and 10 respectively. The error correction terms for both the estimated VEC models are found to be negative and statistically significant, further reinforcing the presence of cointegration among the variables. About 10% of the short run disequilibrium between financial market development, trade openness, GDP per capita and REER gets corrected in the long run. However, only 3.5% of the short run disequilibrium between the broad-based index of financial development, trade openness, GDP per capita and REER gets corrected in the long run. Table 11 indicates that all of the short run coefficients quantifying the impact of financial development on trade openness are statistically significant at the 5% level, except for the three year lag. Table 12 also exhibits a similar short run dynamics.

All the short run coefficients quantifying the impact of financial market development on trade openness are found to be statistically significant at the 5% level, except for the three year lag. Few of the lagged terms of trade openness, real exchange rate and log of GDP per capita are also found to have significant short run impact on trade openness, in case of Indian economy.

Table 9: VEC model results based on Equation (4)

| Variables | Coefficients | P-value | Significance | ECT$_{-1}$(p value)[t ratio] |
|-----------|--------------|---------|--------------|------------------------------|
| $\Delta TRADE_{t-1}$ | 0.8711 | 0.0084 | Yes | -0.035(0.023) [-2.63] |
| $\Delta TRADE_{t-2}$ | -0.3417 | 0.3358 | No | |
| $\Delta TRADE_{t-3}$ | -1.3473 | 0.0018 | Yes | |
| $\Delta TRADE_{t-4}$ | 0.7160 | 0.0512 | No | |
| $\Delta TRADE_{t-5}$ | 0.5232 | 0.0387 | Yes | |
| $\Delta FD_{t-1}$ | -135.7891 | 0.0019 | Yes | |
| $\Delta FD_{t-2}$ | 90.5397 | 0.0125 | Yes | |
| $\Delta FD_{t-3}$ | -50.4864 | 0.3256 | No | |
| $\Delta FD_{t-4}$ | -141.8381 | 0.0026 | Yes | |
| $\Delta FD_{t-5}$ | 166.2704 | 0.0051 | Yes | |
| $\Delta LGDP_{t-1}$ | 165.7287 | 0.0250 | Yes | |
| $\Delta LGDP_{t-2}$ | 145.7564 | 0.0135 | Yes | |
| $\Delta LGDP_{t-3}$ | -64.2459 | 0.1332 | No | |
| $\Delta LGDP_{t-4}$ | 144.3424 | 0.0095 | Yes | |
| $\Delta LGDP_{t-5}$ | -33.1946 | 0.4232 | No | |
| $\Delta REER_{t-1}$ | 0.1477 | 0.3928 | No | |
| $\Delta REER_{t-2}$ | -0.5077 | 0.0137 | Yes | |
| $\Delta REER_{t-3}$ | 1.0760 | 0.0016 | Yes | |
| $\Delta REER_{t-4}$ | -0.1967 | 0.2558 | No | |
| $\Delta REER_{t-5}$ | 0.1963 | 0.2580 | No | |

Source: Authors estimation

Table 10: VEC model results based on Equation (5)

| Variables | Coefficients | P-value | Significance | ECT$_{-1}$(p value)[t ratio] |
|-----------|--------------|---------|--------------|------------------------------|
| $\Delta TRADE_{t-1}$ | 0.8537 | 0.0125 | Yes | -0.1045(0.086) [-1.8824] |
| $\Delta TRADE_{t-2}$ | 0.1158 | 0.6787 | No | |
| $\Delta TRADE_{t-3}$ | -1.1513 | 0.0047 | Yes | |
| $\Delta TRADE_{t-4}$ | 0.6349 | 0.0657 | No | |
| $\Delta TRADE_{t-5}$ | 0.5908 | 0.0483 | Yes | |
| $\Delta FMD_{t-1}$ | -0.6503 | 0.0029 | Yes | |
| $\Delta FMD_{t-2}$ | 66.7695 | 0.0063 | Yes | |
| $\Delta FMD_{t-3}$ | -5.9565 | 0.8087 | No | |
| $\Delta FMD_{t-4}$ | -65.1035 | 0.0121 | Yes | |
| $\Delta FMD_{t-5}$ | 88.5719 | 0.0086 | Yes | |
| $\Delta LGDP_{t-1}$ | 117.6282 | 0.0375 | Yes | |
| $\Delta LGDP_{t-2}$ | 80.1164 | 0.0918 | No | |
| $\Delta LGDP_{t-3}$ | -72.5205 | 0.0748 | No | |
| $\Delta LGDP_{t-4}$ | 90.7226 | 0.0417 | Yes | |
| $\Delta LGDP_{t-5}$ | -39.7234 | 0.3520 | No | |
| $\Delta REER_{t-1}$ | 0.0659 | 0.7053 | No | |
| $\Delta REER_{t-2}$ | -0.4193 | 0.0377 | Yes | |
| $\Delta REER_{t-3}$ | 0.7916 | 0.0037 | Yes | |
| $\Delta REER_{t-4}$ | -0.1862 | 0.2814 | No | |
| $\Delta REER_{t-5}$ | 0.0677 | 0.6793 | No | |

Source: Authors estimation

It is necessary to check the statistical and operational quality of the estimated models, with the help of regression diagnostic tests. Since the study uses financial and macroeconomic variables, it aggravates the chance for collinearity between explanatory variables. However, the significance of regression coefficients and lower variance inflation factor rule out the possibility of high multicollinearity in the model. Therefore, no serious remedial measures are required to treat it. Other tests on residuals ensure that the model is adequately robust and does not suffer from serial autocorrelation, non-normality and heteroscedasticity. The Ramsey RESET (1969) test was conducted to check whether the functional form of model is correct or not. Result shows all three estimated models have correct functional form (see Table A2 in Appendix).

The presence of structural breaks is identified by using sequential Bai-Perron (2003) test and F statistics spotted three structural breaks in the series, results of which are reported in Table A3 of Appendix. Bai Perron Test indicates multiple structural breaks in 1998, 2008 and 2014, and these can be explained as follows: Introduction of second generation banking sector reforms based on Narasimham Committee 2 in 1998, global financial recession in 2008 and the initiation of the fall in the rate of economic growth in 2014 respectively. The validity of model is not affected by the presence of structural breaks because all the macroeconomic variables entering the model are stationary at their first difference based on both the results of traditional unit root tests (ADF and KPSS) and Perron unit root test with structural breaks.

The presence of cointegration among variables hints at the presence of unidirectional or bidirectional Granger causality among variables. Short run causality can be obtained from Chi-square ($\chi^2$) value of lagged difference of independent variables and long run causality can be examined with the help of t statistics on coefficients of lagged values of error correction terms ($ECT_{t-1}$). The results of Granger causality test based on first differenced variables are shown in the table below.

VEC based Granger causality test results reported in Table 11 reject null hypothesis that financial development does not granger cause trade openness, log of GDP per capita does not granger cause trade openness, and real exchange rate does not granger cause trade openness. Hence, unidirectional short run causality is found to run from financial development to trade openness, log of GDP per capita to trade openness and real exchange rate to GDP per capita. Empirical evidence also indicates one way causality from real exchange rate to financial development.

VAR based causality test indicates that at the 5 percent level of significance, there is uni-directional causality in the short run from real exchange rate to trade openness, trade openness to financial institutional development and log of GDP per capita to financial institutional development. Moreover, there is evidence for bidirectional feedback short run causality between financial institutional development and real exchange rate.

Table 11: Granger causality test results based on VECM Equation (4)

| Independent | | | | | |
|---|---|---|---|---|---|
| Dependent | Chi square statistics of lagged First differenced term(p value) | | | | ECT$_{-1}$(P value)[t ratio] |
| | $\Delta$TD | $\Delta$FD | $\Delta$LGDP | $\Delta$REER | |
| $\Delta$ TD | - | 33.10(0.000) | 20.06(0.0012) | 25.98(0.0001) | -0.035(0.023) [-2.63] |
| $\Delta$FD | 5.25(0.38) | - | 7.04(0.219) | 17.71(0.0033) | -0.0002(0.04) [-2.28] |
| $\Delta$LGDP | 4.54(0.46) | 8.83(0.11) | - | 3.20(0.6683) | 0.0001(0.27) [1.13] |
| $\Delta$REER | 1.407(0.92) | 2.92(0.712) | 6.85(0.23) | - | -0.026(0.42) [-0.82] |

Source: Authors estimation. Note: Estimated 5% significance level.

Table 12: Granger causality test results based on VAR

| Independent | | | | |
|---|---|---|---|---|
| Dependent | Chi square statistics of lagged First differenced term(p value) | | | |
| | $\Delta$TD | $\Delta$FID | $\Delta$LGDP | $\Delta$REER |
| $\Delta$ TD | - | 2.89(0.23) | 3.79(0.15) | 6.16(0.04) |
| $\Delta$FID | 7.10(0.027) | - | 8.07(0.017) | 6.90(0.03) |
| $\Delta$LGDP | 0.146(0.929) | 0.118(0.942) | - | 2.867(0.238) |
| $\Delta$REER | 4.67(0.09) | 6.60(0.03) | 3.01(0.22) | - |

Source: Authors estimation

Table 13: Granger causality test results based on VECM Equation (5)

| Independent | | | | | |
|---|---|---|---|---|---|
| Dependent | Chi square statistics of lagged First differenced term(p value) | | | | ECT$_{-1}$(P value)[t ratio] |
| | ΔTD | ΔFMD | ΔLGDP | ΔREER | |
| Δ TD | - | 28.69(0.000) | 15.22(0.009) | 18.1(0.002) | -0.104(0.08) [-1.88] |
| ΔFMD | 3.55(0.61) | - | 7.86(0.16) | 14.55(0.01) | -0.002(0.01) [-3.07] |
| ΔLGDP | 2.54(0.76) | 7.06(0.21) | - | 2.81(0.031) | 0.0003(0.5) [0.69] |
| ΔREER | 2.04(0.84) | 3.2(0.66) | 8.9(0.11) | - | -0.122(0.32) [-1.03] |

Source: Authors estimation

With regard to equation (5), VEC based Granger causality tests results indicate presence of short run unidirectional causality running from financial market development to trade openness, log of GDP per capita to trade openness and real exchange rate to trade openness. Real exchange rate is also found to granger cause financial market development and log of GDP per capita.

**Conclusion and Policy Implications**

This study analyzed whether unidirectional or bidirectional causality relationship exists between financial development and international trade of Indian economy for the time period 1980-2019. Most of the existing studies have examined this macroeconomic relationship by carrying out a cross country analysis with the help of panel data. Indian studies on the subject considered only a narrow measure of financial development based on private credit, and measured international trade performance only for manufactured goods, and excluded trade in agriculture and trade in services. This study fills the above mentioned research gaps and added a new dimension to previous studies by estimating three distinct econometric models using the three indices of financial development created by IMF. Using Johansen cointegration, the study found that there is existence of long run equilibrium relationship between trade openness and the composite index of financial development. It was further established that cointegration also exists between trade openness and index of financial market development. However, no significant long run association was found between trade openness and index of financial institutional development. The direction of causality between the variables of interest is analyzed using VEC based Granger causality test. The causality test results indicate uni-directional causality running from the composite financial development index to trade openness. Uni-directional causality was also found to run from financial market development index to trade openness. There is also evidence of short run uni-directional causality from trade openness to financial institutional development index.

The empirical results indicate that development of financial sector can positively impact upon international trade performance of Indian economy. Since IMF financial development index is composed of sub-indices of financial market development and financial institutional development, the results also implies that financial market development is exerting bigger influence on overall financial development, when it comes to accelerating trade openness. So, it can be interpreted that trading industries are using more external finance or few collateralizable assets. Hence, standalone trade liberalization policies are not advisable for India. Moreover, the trade promotion policies should encompass the productivity and accessibility aspects of external finance to the export oriented industries. India's foreign trade policy should also ensure that access to finance does not become a hurdle for small and medium sized enterprises to enter and flourish in trade related activities. Since financial development can contribute to overall trade openness of India, there is a need for further financial sector reform to further drive economic prosperity of Indian economy.

Trade openness is found to result in the development of financial institutions in the short run, in case of India. As international trade matures, more financial services and insurance will be demanded in response to increased domestic production. It helps in efficient allocation of savings and further expansion of financial system. Theis study underlines the importance of financial development, and specifically policies targeted at development of financial markets, in enhancing the trade openness of Indian economy. Engaging in international trade is also found to promote development of Indian financial institutions, which puts forth a case for greater integration of Indian economy with the world economy.

**Acknowledgment**

The authors are grateful for the valuable feedback received from the participants of the 31[st] Annual Conference of International Trade and Finance Association, held from May 28-29, 2021.

**Bibliography**

Aleksander Aristovnik. 2008. "Short Term Determinants of Current Account Deficit: Evidence from Eastern Europe and the Former Soviet Union." *Eastern European Economics,* 46, no. 1(January): 24-42.

Bas B. Bakker, and Anna Marie Gulde. 2010. "The Credit Bottom in the EU New Member States: Bad Luck or Bad Policies?" *IMF Working Paper,* no. 130 (May): 1-44.

Daniel Perez Liston, and Lawrence McNeil. 2013. 'The Impact of Trade Finance on International Trade: Does Financial Development Matter?' *International Journal Economics and Business Research* 8, no. 1 (January):1-19.

Daria V. Zakharova. 2008. "One-Size-Fits-One: Tailor-Made Fiscal Responses to Capital Flows." *IMF Working Paper*, no. 269 (December):1-27.https://doi.org/10.5089/9781451871272.001

Edward S. Shaw. 1973. "Financial Deepening in Economic Development." *The Journal of Finance* 29, no. 4 (September).

Fatih Yucel. 2009. "Causal Relationships between Financial Development, Trade Openness and Economic Growth: The case of Turkey."*Journal of Social Sciences* 5, no. 1 (January):33-42. 10.3844/jssp.2009.33.42

Fernando Leibovici. 2018 "Financial Development and International Trade" *FRB St. Louis Working Paper*, no. 15 (August): 1-50

Jonas Vlachos, and Helena Svalryd. 2005. "Financial Markets, the Pattern of Specialization and Comparative Advantage. Evidence from OECD countries."*European Economic Review* 49, no. 1 (January):113-144. 10.1016/S0014-2921(03)00030-8

Katsiaryna Svirydzenka. 2016. "Introducing a New Broad-based Index of Financial Development" *IMF Working Paper,* no. 16/5, January.

https://www.imf.org/en/Publications/WP/Issues/2016/12/31/Introducing-a-New-Broad-based-Index-of-Financial-Development-43621

Kenneth Kletzer and Pranab Bardhan. 1987. "Credit markets and patterns of international trade". *Journal of Development Economics* 27, no. 1 (October): 57-70. DOI: 10.1142/S2194565920500232

Kul B. Luintel, and Musahid. 1999. "A Quantitative Reassessment of the FinanceGrowth Nexus: Evidence from a Multivariate VAR."*Journal of Development Economics* 60, no. 2 (December): 381-405.

Leena Ajit Kaushal, and Neha Pathak. 2015. "The Causal Relationship among Economic Growth Financial Development and Trade Openness in Indian Economy."*International Journal of Economic Perspectives* 9, no. 2 (June): 5-22.

Ling Feng, and Ching Yi Lin. 2013. "Financial shocks and Exports."*International Review of Economics and Finance* 26, no.4 (April): 39-55. 10.1016/j.iref.2012.08.007

Michele Cyrille Samba, and Yan Yu. 2009. "Financial Development and International Trade in Manufactures: An Evaluation of the Relation in Some Selected Asian Countries."*International Journal of Business and Management* 12, no. 12 (December):52-69.

Muhammad Shabaz, and Mohammad Mafizur Rahman. 2014. "Exports, Financial Development and Economic Growth in Pakistan."*International Journal of Development Issues* 13, no. 2 (July): 155-170. 10.1108/IJDI-09-2013-0065

Muhsin Kar, Saban Nazlioglu, and Huseyin Agir. 2013. "Trade Openness, Financial Development, and Economic Growth in Turkey: Linear and Nonlinear Causality Analysis."*Proceedings of International Conference of Eurasian Economies, Petersburg, Russia*: 133-143.

Omoke Philip Chimobi. 2010. "The Causal Relationship among Financial Development, Trade Openness and Economic Growth in Nigeria." *International Journal of Economics and Finance 2,* no. 2: 137-147. 10.5539/ijef.v2n2p137

Puneet K. Arora and Jaydeep Mukherjee. 2020. "The nexus between financial development and trade performance. Empirical evidence from India in the presence of endogenous structural breaks". *Journal of Financial Economic Policy* 12, no. 2: 279-291.

Quoy Tan Do, and Andrei A. Levchenko. 2004. "Trade and Financial Development." *World Bank Working Paper*, no. 3347. http://www-wds.worldbank.org/external/default/WDSC ... ered/PDF/wps3347.pdf.

Rajeev Jain, Dhirendra Gajbhiye, and Sousmasree Tewari. 2019. "Cross Border Trade Credit: A post Crisis Empirical Analysis for India." *RBI Working Paper,* no. 02.https://rbidocs.rbi.org.in/rdocs/Publications/PDFs/WPS022019CBTCCC06C3A2CF6F4 26EBA4590C599B38369.PDF.

Raymond W. Goldsmith. 1969. Financial Structure and Development. New Haven, CT: Yale University Press.

RBI (Reserve Bank of India). 2007. "Report on Currency and Finance." May 31. https://www.rbi.org.in/scripts/PublicationsView.aspx?id=9237

Richard E. Baldwin. 1989. Exporting the capital markets: comparative advantages and capital markets imperfections. North Holland: The convergence of international and domestic markets: Amsterdam. 135-152.

Ronald I. McKinnon. 1973. Money and Capital in Economic Development. Washington, D.C: Brookings Institution Press.

Ross Levine, and Sarah Zervos. 1998. "Stock Markets, Banks, and Economic Growth." *American Economic Review* 88, no. 3 (June): 537-558.

Ross Levine, Norman Loayza, and Thorsten Beck. 2000. "Financial Intermediation and Growth: Causality and causes."*Journal of Monetary Economics* 46, no. 1 (August):31-77.

Ross Levine. 1997. "Financial Development and Economic Growth: Views and Agenda."*Journal of Economic Literature* 35, no. 2 (June): 688-726.

Thorsten Beck. 2002. "Financial Development and International Trade, Is there a link?" *Journal of International Economics* 57, no. 1 (June): 107-131.

Victor Murinde, and Fern S.H. Eng. 1994. "Financial development and economic growth in Singapore: demand following or supply leading?"*Applied Financial Economics* 4, no. 6: 391–404.

Vlatka Bilas, Mile Bosnjak, and Ivan Novak. 2017. "Examining the Relationship between Financial Development and International Trade in Croatia." *South East European Journal of Economics and Business* 12, no. 1(July): 80-88. https://doi.org/10.1515/jeb-2017-0009

World Bank. 2021. *World Development Indicators*. Washington D.C.

Youssouf Kiendrebeogo. 2012. "Understanding the Causal Links between Financial Development and International Trade." Series Etudes et Documents Du Cerdi, no. 34, (September).

# Appendix

Table A1: ADF and KPSS Stationarity test Results

| Deterministic terms | Variables | Levels | | | 1st Differences | | |
|---|---|---|---|---|---|---|---|
| | | Test statistic | Critical value | Remarks | Test statistic | Critical value | Remarks |
| | ADF Unit root test | | | | | | |
| Intercept | TRADE | -0.66 | -2.94 | I(1) | -5.19 | -2.94 | I(0) |
| | FD | -1.45 | -2.94 | I(1) | -4.72 | -2.94 | I(0) |
| | FID | 0.24 | -2.94 | I(1) | -4.95 | -2.94 | I(0) |
| | FMD | -1.87 | -2.94 | I(1) | -4.68 | -2.94 | I(0) |
| | LGDP | 3.50 | -2.94 | I(1) | -4.73 | -2.94 | I(0) |
| | REER | -2.65 | -2.94 | I(1) | -4.91 | -2.94 | I(0) |
| Intercept & Trend | TRADE | -1.52 | -3.53 | I(1) | -5.11 | -3.53 | I(0) |
| | FD | -2.52 | -3.53 | I(1) | -4.67 | -3.53 | I(0) |
| | FID | -1.17 | -3.53 | I(1) | -5.04 | -3.53 | I(0) |
| | FMD | -2.37 | -3.53 | I(1) | -4.65 | -3.53 | I(0) |
| | LGDP | -0.82 | -3.53 | I(1) | -6.33 | -3.53 | I(0) |
| | REER | -0.81 | -3.53 | I(1) | -6.24 | -3.53 | I(0) |
| | KPSS Unit root test | | | | | | |
| Intercept | TRADE | 0.65 | 0.46 | I(0) | 0.14 | 0.46 | I(1) |
| | FD | 0.59 | 0.46 | I(0) | 0.06 | 0.46 | I(1) |
| | FID | 0.58 | 0.46 | I(0) | 0.21 | 0.46 | I(1) |
| | FMD | 0.42 | 0.34*** | I(0) | 0.11 | 0.46 | I(1) |
| | LGDP | 0.75 | 0.46 | I(0) | 0.735 | 0.739* | I(1) |
| | REER | 0.46 | 0.34*** | I(0) | 0.46 | 0.73* | I(1) |

| Intercept & Trend | TRADE | 0.32 | 0.14 | I(0) | 0.13 | 0.14 | I(1) |
|---|---|---|---|---|---|---|---|
| | FD | 0.18 | 0.14 | I(0) | 0.05 | 0.14 | I(1) |
| | FID | 0.16 | 0.14 | I(0) | 0.09 | 0.14 | I(1) |
| | FMD | 0.12 | 0.11*** | I(0) | 0.06 | 0.14 | I(1) |
| | LGDP | 0.20 | 0.14 | I(0) | 0.02 | 0.14 | I(1) |
| | REER | 0.19 | 0.14 | I(0) | 0.08 | 0.14 | I(1) |

Source: Authors estimation. Note: Unit root tests are carried out at 5% significance level. *indicates 1% level of significance and *** indicates 10% level of significance. The rest of unit root tests are given at 5% significance level. I(0) means integrated of order zero and I(1) means integrated of order one. The lag length is selected using Schwarz Info Criterion

Table A2: Regression Diagnostic Test Results

| Equation 4 | | | Equations 6, 7, 8 and 9 | | Equation 5 | |
|---|---|---|---|---|---|---|
| **Diagnostic Tests** | **P Value** | **Result Description** | **P Value** | **Result Description** | **P Value** | **Result Description** |
| a) Breusch Godfrey LM<br>Lags<br>1.<br>2.<br>3.<br>4. | <br><br>0.36<br>0.27<br>0.67<br>0.86 | There is no autocorrelation | <br><br>0.68<br>0.63<br>0.07<br>0.48 | There is no autocorrelation | <br><br>0.62<br>0.26<br>0.94<br>0.18 | There is no autocorrelation |
| b) Jaque-Bera<br>• Joint | 0.71 | Residuals are normally distributed | 0.83 | Residuals are normally distributed | 0.80 | Residuals are normally distributed |
| c) White heteroscedasticity<br>• Chi-sq | <br><br>0.36 | We do not have heteroscedasticity | <br><br>0.30 | We do not have heteroscedasticity | <br><br>0.422 | We do not have heteroscedasticity |
| d) Ramsey RESET<br>• t-statistic<br>• F-statistic<br>• Likelihood Ratio | <br>0.10<br>0.10<br><br>0.08 | Our model is correctly specified | <br>0.58<br>0.58<br><br>0.56 | Our model is correctly specified | <br>0.06<br>0.06<br><br>0.051 | Our model is correctly specified |
| e) Variance Inflation Factors<br>• FD<br>• LGDP<br>• REER | <br><br>2.58<br>3.06<br>1.56 | No Serious Multicollinearity is present (VIF<10) | <br><br>8.94<br>7.54<br>5.11 | No Serious Multicollinearity is present (VIF<10) | <br><br>1.60<br>1.72<br>1.76 | No Serious Multicollinearity is present (VIF<10) |

Source: Authors estimation

Table A3: Sequential Bai-Perron test results

| Equation(1) | | | Equation(2) | | Equation(3) | |
|---|---|---|---|---|---|---|
| **Break test** | **F-statistic** | **Critical value** | **F-statistic** | **Critical value** | **F-statistic** | **Critical value** |
| 0 vs. 1* | 39.20 | 16.19 | 37.59 | 16.19 | 36.16 | 16.19 |
| 1 vs. 2* | 27.31 | 18.11 | 19.99 | 18.11 | 19.73 | 18.11 |
| 2 vs. 3* | 19.15 | 18.93 | 20.20 | 18.93 | 25.67 | 18.93 |
| 3 vs. 4 | 5.4 | 19.64 | 11.86 | 19.64 | 13.77 | 19.64 |

Source: Authors estimation          *Significant at the 0.05 level.